\documentclass[runningheads]{llncs}
\usepackage[T1]{fontenc}
\usepackage{graphicx}

\usepackage{enumitem}
\usepackage[super]{nth}
\usepackage{multirow}
\usepackage{algorithm}
\usepackage[noend]{algorithmic}
\usepackage{booktabs}
\usepackage{soul} 
\usepackage{comment}
\usepackage{enumitem}

\usepackage{url}
\newcommand{\hashtag}[1]{\texttt{\##1}}
\newcommand{\mention}[1]{\texttt{@#1}}
\usepackage{xcolor}

\begin{document}
\title{Temporal Nuances of Coordination Network Semantics}
%
%
\author{Derek Weber\inst{1,2}\orcidID{0000-0003-3830-9014} \and
Lucia Falzon\inst{3,4}\orcidID{0000-0003-3134-4351}}
\authorrunning{D.C. Weber and L. Falzon}
%
\institute{School of Computer Science, University of Adelaide, South Australia, Australia \and
Defence Science and Technology Group, Edinburgh, South Australia, Australia
\email{derek.weber@\{adelaide.edu.au,defence.gov.au\}}
\and
School of Psychology, University of Melbourne, Victoria, Australia\\
\email{lucia.falzon@unimelb.edu.au}
\and
School of Mathematical Science, University of Adelaide, South Australia, Australia }
\maketitle              
\begin{abstract}
Current network-based methods for detecting coordinated inauthentic behaviour on social media focus primarily on inferring links between accounts based on common `behavioural traces' \cite{Pacheco2021icwsm}, such as retweeting the same tweet or posting the same URL. Assuming the goal of coordination is amplification, boosting a message within a constrained period, most approaches use a temporal window to ensure the \emph{co-activity} occurs within a specific timeframe 
\cite{graham2020virus,magelinski2022,Pacheco2021icwsm,weber2021coord}. 
Real-world application requires considering near real-time processing, creating performance requirements, which also highlight gaps in the semantics of coordination in the literature. 
These methods could all exploit temporal elements of coordinated activity. 
We describe preliminary research 
regarding coordination network semantics,
coordination network construction, relevant 
observations in three political Twitter datasets
and the role of cheerleaders in revealing social bots.

\keywords{Coordination Networks \and Social Media \and Social Network Analysis.}
\end{abstract}
%
%
%
\section{Introduction}\label{sec:intro}



There is ongoing evidence of threats to our \emph{social cybersecurity} \cite{Carley2020} in the form of accounts working together to spread narratives as part of overt activism and political campaigning or covert, potentially malicious attempts to spread mis- and disinformation and influence society, including across international borders 
\cite{Dawson2019,graham2020virus}.
Much of this coordination relies on amplifying, or \textit{boost}ing \cite{weber2021coord}, a preferred narrative on social media, though coordination can be also used to exacerbate abuse 
\cite{graphika2020,PachecoFM2020whitehelmets,ratkiewicz2011}. 
Research into these kinds of cooperative efforts fall into two categories: 1) campaign detection, emerging from spam detection in Web 1.0 email, and Web 2.0 social media (e.g., \cite{LeeCCS2013campext}); and more recently, 2) detection of coordinating groups of social media accounts (e.g., 
\cite{graham2020virus,magelinski2022,Nizzoli2021,Pacheco2021icwsm,weber2021coord}). 
Forensic efforts enable deeper understanding, especially regarding the information cycle including the mainstream media, but these require more manual examination (e.g., 
\cite{Benkler2018,Dawson2019,Jamieson2020,SingerB2019likewar}).

\begin{figure}[ht!]
    \centering
    \includegraphics[width=0.99\columnwidth]{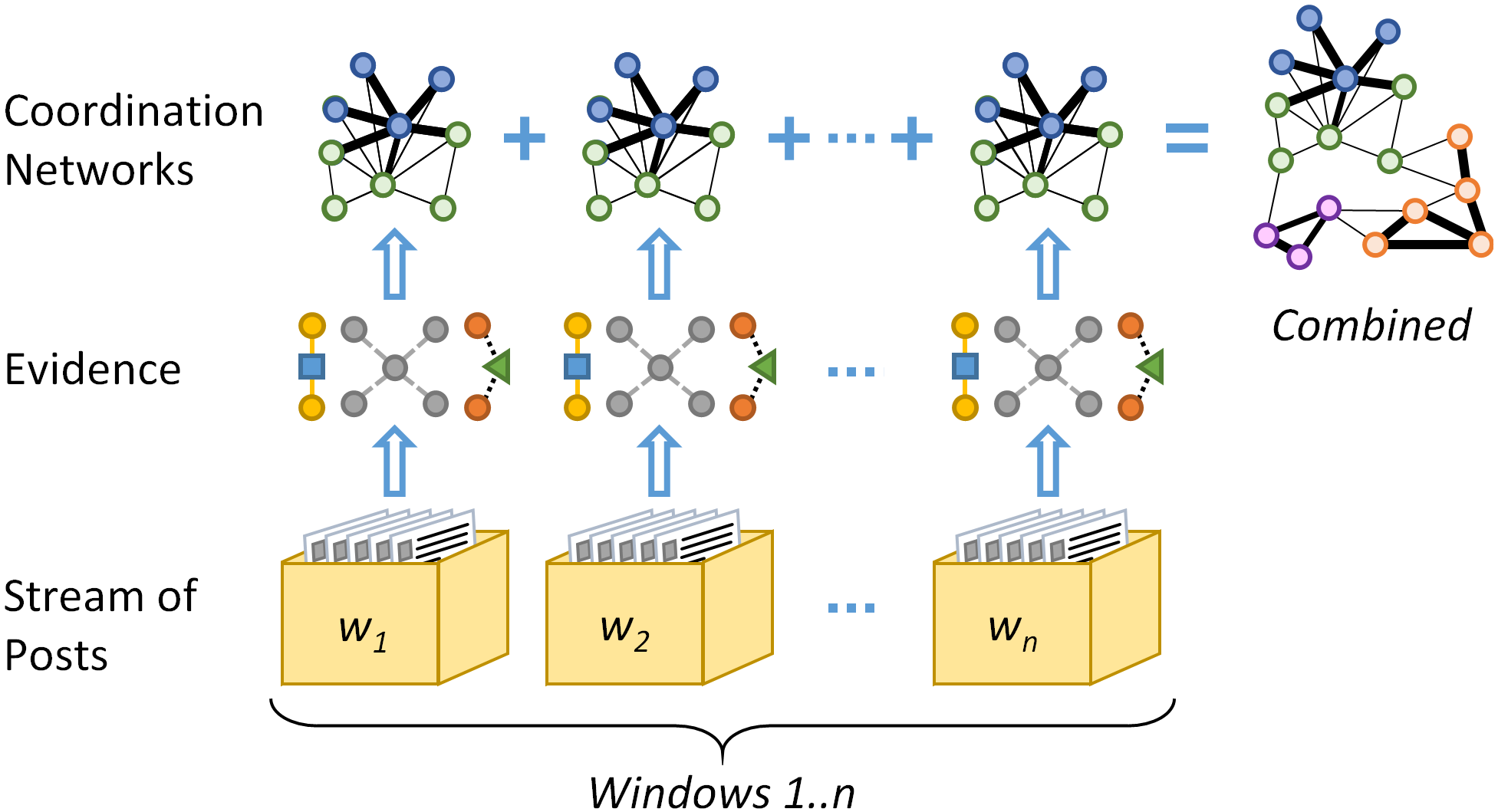}
    \caption{Construction of a synchronised action network. A stream of social media posts is broken into fixed windows (bottom), and each of these is inspected for evidence of various types of coordination, from which a weighted undirected account network is built. These are then combined in post-processing and mined for \emph{highly coordinating communities}.}
    \label{fig:san_cons}
\end{figure}

Many of the current coordination detection techniques rely on finding anomalously high levels of 
concurrent 
behaviour, 
such as retweeting the same tweet, using the same 
hashtag, URL, or 
text (especially across social media platforms). We refer to these as \emph{co-activities} \cite{ratkiewicz2011,weber2021coord}. Such methods exploit the temporal aspect of social media activity for two reasons:
\begin{enumerate}
    \item To constrain coincidental activities to a specific time window (e.g., examine only posts within a $10$ second time window), with small windows indicating inauthentic behaviour \cite{magelinski2022,PachecoFM2020whitehelmets,weber2021coord}; and
    \item to reduce computational load by only considering activity within the time window \cite{magelinski2022,weber2021coord}.
\end{enumerate}
All approaches in the literature build undirected weighted networks of accounts linked according to evidence of coordination, either across entire social media datasets, which can only be performed post-collection, or via aggregation of results from individual windows, which could be calculated progressively (e.g., see Figure~\ref{fig:san_cons}).
To the authors' knowledge, there has been no discussion of the 
finer-grained temporal analysis of the coordinated activities
themselves, 
which are lost in the construction of such coordination networks. Tracking the evolution of the ties in \emph{highly coordinating communities} (HCCs)\footnote{Community extraction methods identifying HCCs are intended to avoid the common coincidental behaviour that otherwise may appear coordinated.} within these networks, and their semantics, may reveal further insights into information campaigns. 
This will also be required for ongoing or progressive analysis in any real-time or near real-time applications.

The perfect real-time solution would require searching for relevant time-bounded co-activities on the arrival of every new social media post, creating new edges in the coordination network as matches are found. This approach creates significant performance costs, as it is not the entire coordination network that is always of interest, but of the HCCs within. A more relaxed approach involving discrete or overlapping windows (i.e., review new posts in batches every $x$ seconds) will delay the discovery of some coordination and potentially miss coordination spanning window boundaries, but at a much lower computational cost \cite{weber2021coord}. 

This paper explores issues raised while tracking the evolution of HCCs and considers the semantics of coordination links and their aggregation, and in doing so identifies the phenomenon of cheerleader coordination through relevant experiences from social media analyses. 
Associated code and data can be found at \url{https://github.com/weberdc/find_hccs}.




\section{The State of the Art of Coordination Detection}\label{sec:coord}

Since at least $2019$, attempts have been made to create general coordination detection frameworks 
\cite{magelinski2022,Nizzoli2021,Pacheco2021icwsm,weber2021coord} 
which reveal frequent indirect connections between accounts based on common actions. These include retweeting the same tweet 
\cite{Keller2019,Pacheco2021icwsm,weber2021coord}, 
using the same URL
\cite{CaoCLGC2015urlsh,magelinski2022,ratkiewicz2011,weber2021coord}, 
similar text 
\cite{graham2020virus}, 
and similar media \cite{Pacheco2021icwsm,Yu2021}, 
and any other `behavioural trace' \cite{Pacheco2021icwsm} that could link accounts (e.g., swapping account names 
\cite{Pacheco2021icwsm} or activity patterns 
\cite{Dawson2019}). 
Coordination networks of accounts linked by edges weighted by the evidence linking them are the result, referred to variously as `user similarity networks' \cite{Nizzoli2021}, `latent coordination networks' \cite{weber2021coord}, or, when explicitly referring to the temporally constrained nature of the coordination, `synchronized action networks' \cite{magelinski2022}. HCCs within these coordination networks may be simple connected components \cite{Pacheco2021icwsm}, or obtained by filtering edges by weight \cite{Pacheco2021icwsm,weber2021coord} or cluster detection methods (e.g., $kNN$ \cite{CaoCLGC2015urlsh}, 
FSA\_V \cite{weber2021coord}, multi-resolution filtering \cite{Nizzoli2021} and multi-view modularity \cite{magelinski2022}). While many approaches in the literature are validated through case studies and examples, a range of computational validation techniques also exist confirming the value of a network-based approach \cite{weber2021coord}.


Some detection methods disregard common behaviour types and use ``synchronicity analysis''~\cite[p.~251]{Dawson2019}, through the application of sliding windows to detect contemporaneous co-activities. 
Weber \& Neumann use adjacent discrete windows, aggregating coordination networks built from each and then mining the result for HCCs \cite{weber2021coord}, while Graham et al. overlap discrete windows to avoid missing evidence 
\cite{graham2020virus}. 
Pacheco et al. \cite{Pacheco2021icwsm} and Dawson \& Innes \cite{Dawson2019} use fully overlapping windows without providing detailed descriptions, and Pacheco et al. and Magelinski et al. \cite{magelinski2022} discuss the balance between window size and false positives, while Weber \& Neumann explore it systematically \cite{weber2021coord}. 

Regardless of whether the coordination networks are constructed through post-collection analysis or aggregated through near real-time window-based analysis, the removal of the temporal element may lead to invalid associations between actors in the networks. 
In the same way people have different but separate social circles that they may meet with at different times \cite{simmel1908geheimnis}, if an account, X, actively interacts with one set of accounts early in the dataset, but then a different set of accounts late in the dataset, there may be no real connection between those sets of accounts except for X. Aggregating these groups in post-collection analysis may give the impression of a single large community, however. That said, interactions with different social circles may be interleaved also (e.g., chatting with work colleagues during the work day, but family and friends every evening), so simply grouping accounts into social circles based on whether the interactions appear early or late in the dataset is also insufficient. Coordination via regular but temporally distant interactions will be of less use for coordinated amplification, however, which relies on bursts of common activity \cite{weber2021coord}. 

\subsection{Clarification of Terminology}

Due to variations in terminology in the literature, we clarify definitions here.
\begin{description}
    \item[Action type (\textit{cf.} \cite{magelinski2022})] The type of interactions 
    that a user employs while using social media (e.g., reposting, using a URL or mention). These can be combined as higher order action types \cite{magelinski2022}, and used together in ``coordination strategies'' \cite[p.~1]{weber2021coord}. Examples include Morstatter et al.'s study of retweets and mentions during the $2017$ German election \cite{morstatter2018alt} and Schliebs et al.'s examination of retweets and replies in their study of coordinated amplification of foreign propaganda in the UK \cite{schliebs2021}.
    \item[Coordination action (\textit{co-action})] An instance of an action type which `coordinates' with another co-action. Also referred to as a \textit{co-activity}.
    \item[Reason] The value that links two accounts by a coordination action, e.g., the URL, hashtag, or account they both use.
    \item[Time window] A period of time with a duration of $\gamma$ units (e.g., minutes, seconds). Windows may overlap (e.g.,~\cite{graham2020virus,magelinski2022}) or may be adjacent (e.g.,~\cite{weber2021coord}).
    \item[Coordination] A pair of co-actions occurring within a given time window. N.B., as will become apparent, this definition overlooks the intent of the participants, as accounts may coordinate unintentionally using this definition. 
    \item[Coordination network (CN)] A weighted network of accounts, linked based on evidence of their coordination. 
    \item[Highly coordinating community (HCC)] A subset of a coordination network of accounts extracted systematically to avoid incidental coordination.
\end{description}
\section{The Challenge}\label{sec:challenge}



We first provide an example demonstrating the issue of combining temporally distant coordination networks and how deeper temporal analysis can reveal `cheerleaders' that bind unwitting accounts, calling into question how the term `coordination' should be used. We then provide a systematic consideration of the variations of window overlap and action type.

\subsection{Spurious Associations and Cheerleaders}

\begin{figure}[ht]
    \centering
    \includegraphics[width=0.64\columnwidth]{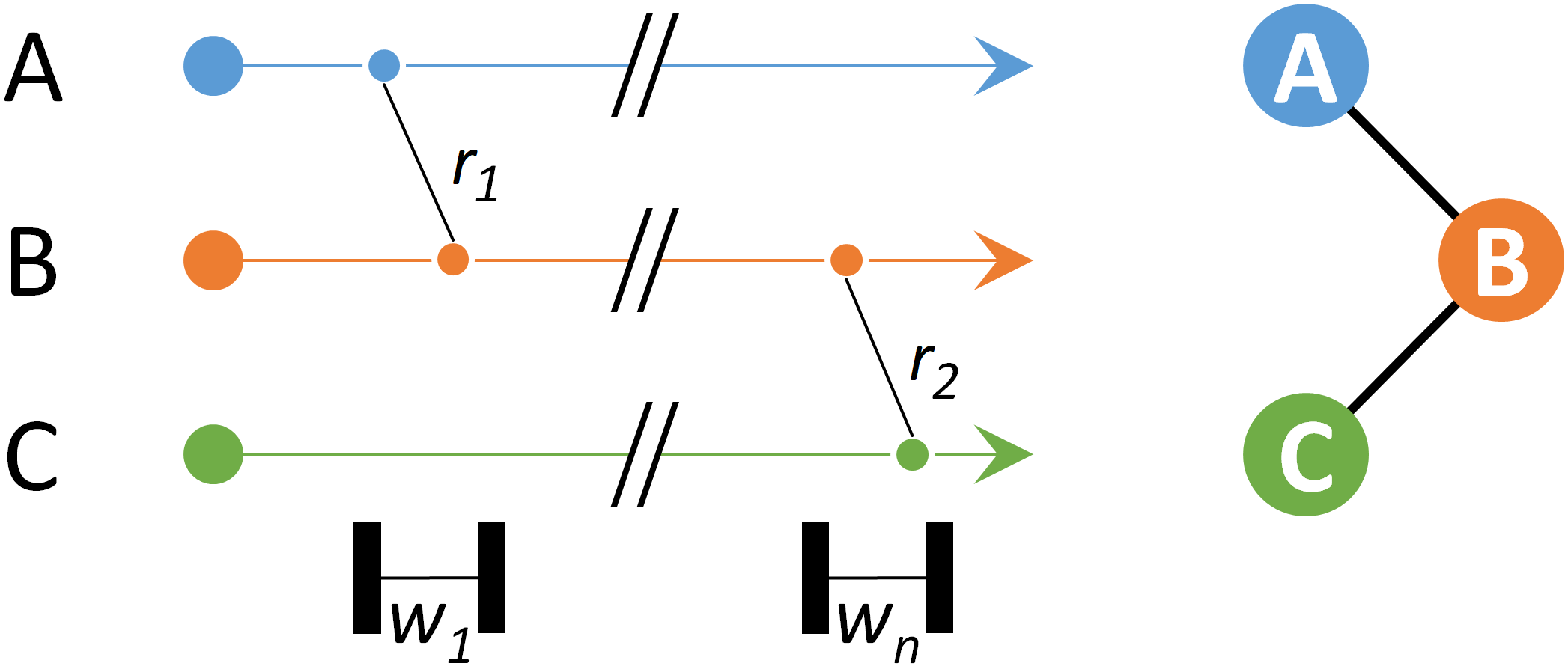}
    \caption{Timelines for accounts $A$, $B$, and $C$. $A$ and $B$ coordinate somehow (based on reason $r_1$) in time window $w_1$, while $B$ and $C$ coordinate in time window $w_n$ (based on reason $r_2$). When the coordination networks for each window are combined (i.e., aggregated, right), the resulting transitive association between $A$ and $C$ may be spurious.}
    \label{fig:ab_bc_separate}
\end{figure}

Ignoring the temporal distance between coordination actions may result in erroneously creating communities whose members are entirely unrelated. 
Consider the situation where accounts $A$ and $B$ coordinate in window $w_1$ for reason $r_1$, and then much later accounts $B$ and $C$ coordinate in window $w_n$ for reason $r_2$ (Figure~\ref{fig:ab_bc_separate}). In this simple example the post-collection combination of each window's coordination networks would result in a network $G=(V,E)$ with nodes $V=\{A,B,C\}$ and edges $E=\{A-B,B-C\}$ even if $r_1 \neq r_2$. This arrangement implies some kind of relationship between $A$ and $C$ (they are members of the same community). This does not hold in the general case, but depends on the nature of the coordination. For example, if $A$ and $B$ both mention URL $u_1$ and then $B$ and $C$ mention the entirely different URL $u_2$, then there is no reason to associate $A$ and $C$, except that links are inferred between both of them and $B$, yet they are still members of the same CN. Simple filtering by edge weight may also result in the removal of $B$, which may be a genuine coordinating account. 



We may identify the introduction of spurious associations by examining the relative timing of pairs of co-actions. If $B$'s co-action always occurs after $A$'s (but within the nominated time window), then it can be argued that $B$ is reacting to $A$, acting as a \emph{cheerleader}, which will be detected even if $A$ is an unwitting participant. If $B$ acts as a cheerleader for many accounts, forming an HCC around it, this recurring pattern of posting second will become apparent and can contribute to star-shaped HCCs, informing our interpretation of the HCC. 

\begin{figure}[ht]
    \centering
    \includegraphics[width=0.99\columnwidth]{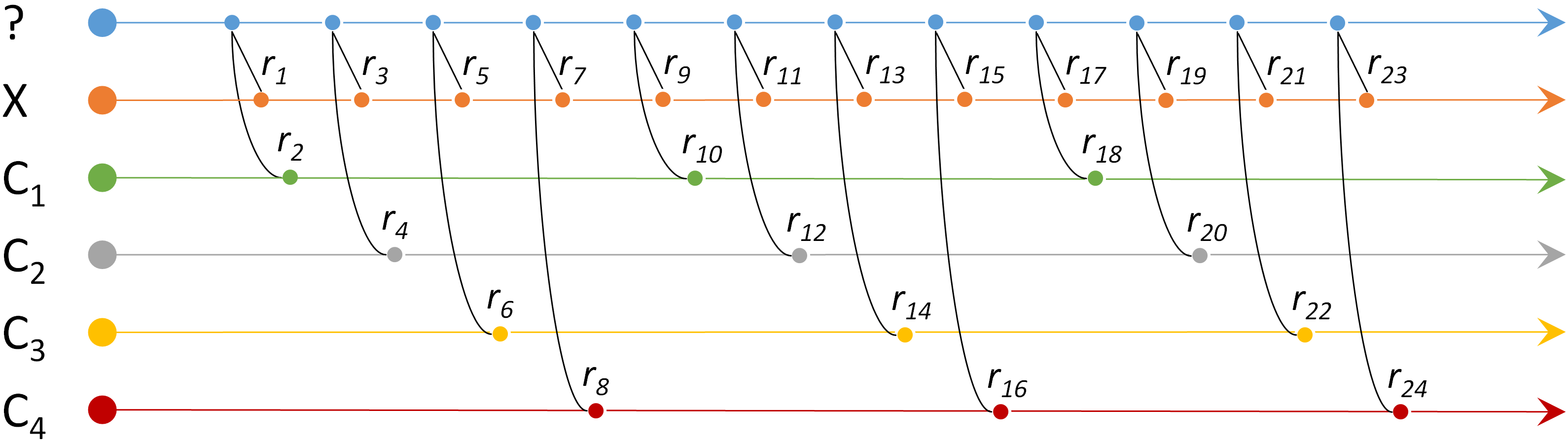}
    \caption{A team of cheerleaders boost any content promoted by X (e.g., via retweet). Each tweet retweeted by X (via $r_1$ to $r_{23}$) is also retweeted by one of the cheerleaders, e.g., $r_1$ and $r_2$ are retweets of the same tweet. }
    \label{fig:cheerleader_example}
\end{figure}

Coordinated cheerleading of an account may produce a star-shaped HCC if the cheerleaders are attempting to hide their activities by avoiding the same co-actions. Figure~\ref{fig:cheerleader_example} shows a situation where the target of cheering, X, regularly retweets an original tweet (represented by a post by an arbitrary user, `?').\footnote{The example could easily be extended to X mentioning a hashtag or URL, as long as the time window considered does not capture more than one cheerleader.} Each time X does this, one of four cheerleaders ($C_1$, $C_2$, $C_3$, and $C_4$) performs the corresponding co-action within a given time window in a \textit{round robin} fashion. In this way, the cheerleaders amplify what X is posting but avoid strong associations to each other. The resulting HCC will include edges between the cheerleaders and X (each with a weight of $3$) but not directly between the cheerleaders. Cheerleading campaigns will balance the time between their activities based on the level of activity in the general discussion, so their co-actions are considered \emph{coincidental}.

\subsection{Strengthening HCC Ties using the Temporal Element}


When analysing identified HCCs, we can use the temporal element of coordination to make explicit edges that are otherwise missed due to time window boundaries, and simultaneously identify missing associations between HCC members. Doing this depends on two variables: 1) the nature of coordination under consideration; and 2) the relative proximity of the two time windows. For co-actions that rely on unique conditions, e.g., retweeting a tweet (which usually occurs only once), some combinations of factors are impossible, such as $A$ and $B$ co-retweeting a tweet in one window, and then $B$ and $C$ co-retweeting the same tweet in a later window. For most action types (e.g., using the same hashtag, mention or URL) this restriction does not apply.


So far, to build CNs, accounts are associated in pairs based on action type and coordination reason, and then the pairs are 
aggregated as ties in a social network of accounts.
An extension to this is to consider reasons and temporal distance when 
forming the ties in this network.

\begin{figure}[ht]
    \centering
    \includegraphics[width=0.99\columnwidth]{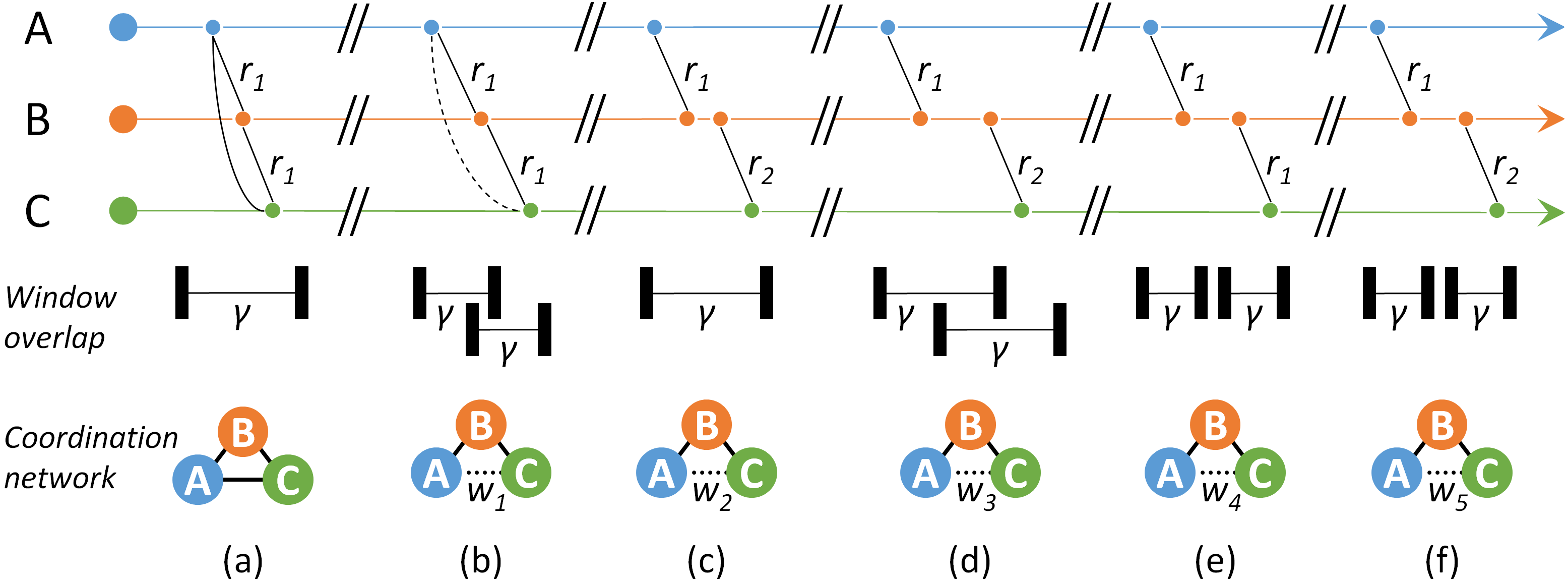}
    \caption{Timelines for accounts $A$, $B$, and $C$ whose behaviour is coordinated due to reasons $r_1$ and $r_2$ in independent examples (a) to (f), within time windows of $\gamma$ duration. Edge weights in each example's CN are $1$, except where labelled $w_1$ to $w_5$. }
    \label{fig:ab_bc_all}
\end{figure}

Now we consider the degree of overlap between time windows and how that relates to the reason for coordination. Figure~\ref{fig:ab_bc_all} shows six examples where $A$ and $C$ could be connected by an edge $A-C$ with a weight $w \in [0,1]$, depending on the reason for connecting with $B$ ($r_1$ or $r_2$) and the overlap in time windows $\gamma$ wide. 
For example:
\begin{enumerate}[label={(\alph*)}] 
    \item $A$ and $C$ are both linked to $B$ for reason $r_1$ within the same time window, and thus also directly coordinate with each other, and so the resulting coordination network has an edge $A-C$ with weight $1$. 
    \item $A$ and $C$ could be linked with weight $w_1$ given they both coordinate with $B$ for the same reason $r_1$, and are only prevented from being directly linked because the gap in time between their actions is greater than $\gamma$.
    \item $A$ and $C$ are linked to $B$ for different reasons, $r_1$ and $r_2$, but there is an implied connection because of the temporal proximity of their associations with $B$, and thus a potential link could have weight $w_2$.
    \item $A$ and $C$ are both again linked to $B$ for different reasons, but within overlapping windows, suggesting that there may be an association, but with a weaker weight $w_3 \leq w_2$.
    \item $A$ and $C$ both coordinate with $B$ in separate time windows, which is the situation in Figure~\ref{fig:ab_bc_separate}, and there is an implied transitive connection between $A$ and $C$ with weight $w_4$. Given $r_1 = r_2$, there is reason to set $w_4 > 0$, and its value could be time-sensitive, degrading proportional to the temporal distance (\emph{cf.} the t-SED measure \cite{Kim2019}).
    \item $A$ and $C$, again, both coordinate with $B$ in separate time windows. As $r_1 \neq r_2$, $w_5$ could be set to $0$, as they are only linked by $B$, or it could be set based on the similarity of $r_1$ and $r_2$ and the temporal distance between the co-actions.
\end{enumerate}
Intuitively, the strength of potential association between $A$ and $C$ decreases from the first to the last of these examples, i.e.,  $1 \geq w_1 \geq w_2 \geq w_3 \geq w_4 \geq w_5 \geq 0$.


By incorporating non-zero weights $w_1$ to $w_5$, CNs may consolidate into HCCs that better reflect how genuine the coordination is. This could aid community extraction methods that rely on HCC density (e.g., the multi-view modularity clustering technique used by \cite{magelinski2022}). Application in near real time settings could constrain how far 
back 
to look for coordination, permitting a second broader sliding window, $k\gamma$ units long, for some constant or variable factor $k$ (\emph{cf.} \cite{AssenmacherATG2020,Kim2019,weber2021coord}).


Another approach to a creating a CN with a single connected component based on sparse coordination (i.e., Figure~\ref{fig:ab_bc_separate}) is to create a multi-layer network (\textit{cf.}~\cite{magelinski2022}) in which each layer represents the coordination from a time window, and inter-layer edges connect the same accounts in each layer, the weight of which could be inversely proportional to how far apart the windows are (\emph{cf.} \cite{Kim2019} again). This would expose opportunities to use layer centrality measures, and edge overlap correlations to identify recurrent coordination, but alone does little to inform us regarding spuriously associated accounts. 
Recurrent coordination has been previously highlighted by incorporating behaviour from past windows with a decay factor~\cite{weber2021coord}.
\section{Experiences in Analysis}\label{sec:experiences}

Examples from a number of co-retweeting studies illustrate the points we have been discussing. These examined $120$k tweets by $20$k accounts from a 2018 Australian election (DS1) and the Internet Research Agency's (IRA) $1.5$m tweets by $1.3$k accounts during 2016 (DS2), plus a novel Twitter dataset from the 2020 Republican National Convention (RNC), all described in \cite{weber2021coord} (we used their data). The RNC dataset has $1.5$m tweets over $4$ days by $441.4$k accounts, including $1.2$m retweets by $50$k unique accounts.\footnote{Data were managed according to ethics protocol H-2018-045 (University of Adelaide).}

\subsection{Star-shaped HCCs}
\begin{figure}[pht]
    \centering
    \includegraphics[width=0.99\columnwidth]{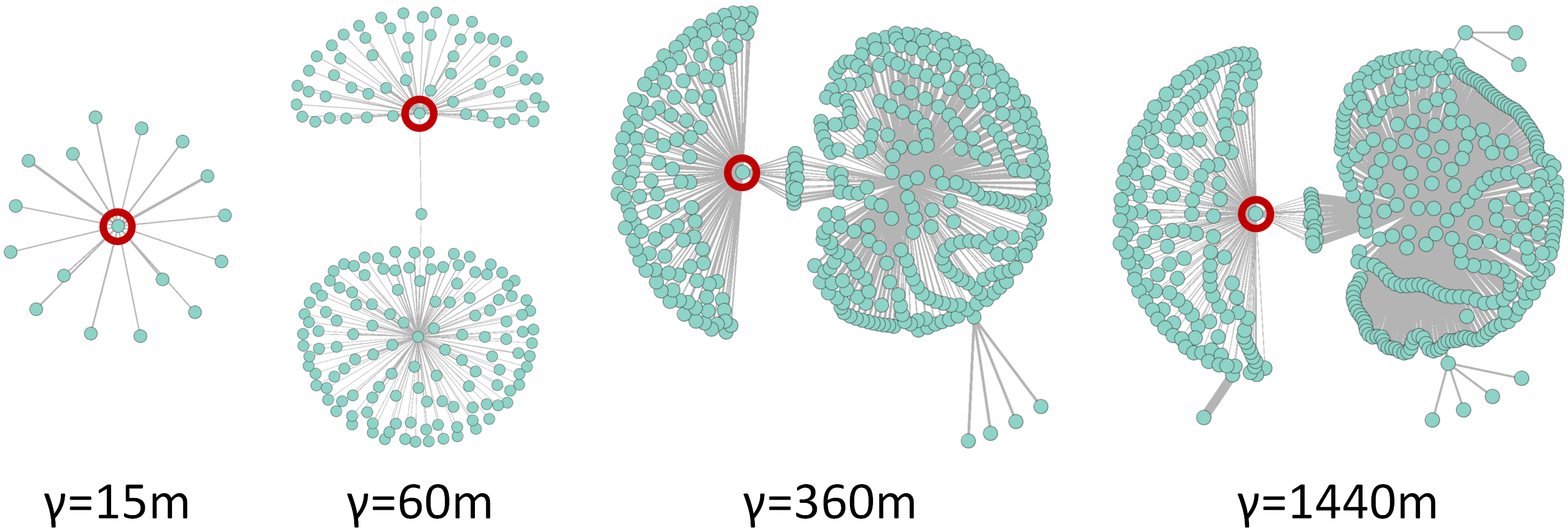}
    \caption{A co-retweeting HCC focused on a specific account (circled) from DS1 \cite{weber2021coord}, which grew as $\gamma$ increased. Visualised with \emph{visone} (\protect\url{https://visone.info}). }
    \label{fig:ds1_hcc_growth}
\end{figure}

\begin{figure}[pth]
    \centering
    \includegraphics[width=0.85\textwidth]{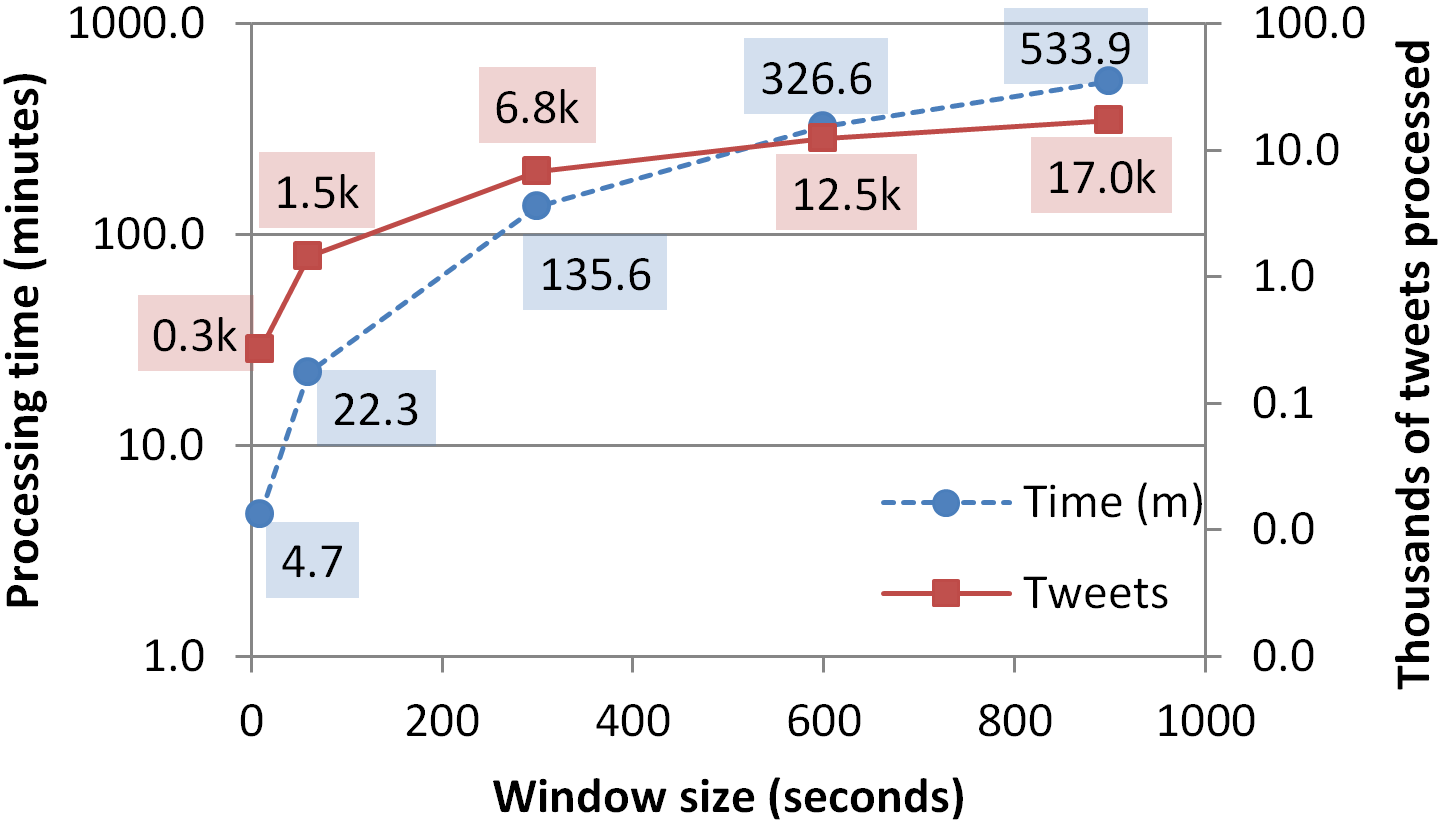}
    \caption{Execution time to identify co-retweets in the RNC dataset in different time windows (using log scales).}\label{fig:runtimes_vs_tweets}
\end{figure}

\sloppy The DS1 study revealed star-shaped HCCs at different time windows ($\gamma$=$\{15,60,360,1440\}$ minutes), which grew consistently as $\gamma$ increased (Figure~\ref{fig:ds1_hcc_growth}). Although this collection was centred on an Australian election, the use of \hashtag{liberals} (the name of an Australian political party) drew in several American Trump supporters, all linked by co-retweeting one tweet $191$ times and centred around the circled account. Although it is clear that the longer time windows allowed for larger HCCs, it is not clear why the star shapes are so pronounced, given it is unlikely that these accounts were \emph{all} engaging in cheerleading (as in Figure~\ref{fig:cheerleader_example}). It is possible, however, that the central node was a super-retweeter, and others coincidentally retweeted the same tweets. The longer the time window chosen, the more co-activities are captured, but the more filtering is required to remove accidental coordination, leading to exponentially greater processing demands. Even just identifying co-activities quickly becomes expensive, due to the pairwise processing requirements \cite{weber2021coord}, as can be seen in Table~\ref{tab:running_costs} and plotted in Figure~\ref{fig:runtimes_vs_tweets}. Although the relationship between window size and processing time is close to linear, it is clear that processing time gradually increases more quickly as the number of tweets increases.

\begin{table}[pth]
    \centering
    \caption{Execution time to identify co-retweets in the RNC dataset in different time windows. `m' = minute, `s' = second.}
    \label{tab:running_costs}
        \begin{tabular}{@{}rrr@{}}
            \toprule
            Window & Time & Largest batch \\
            \multicolumn{1}{c}{($\gamma$)} & (m:s) & \multicolumn{1}{c}{(tweets)} \\
            \cmidrule{1-1} \cmidrule(l){2-3}
            10s &    4:43 &    268 \\
             1m &   22:16 &  1,453 \\
             5m &  135:33 &  6,824 \\
            10m &  326:33 & 12,459 \\
            15m &  533:54 & 17,019 \\
            \bottomrule
        \end{tabular}
\end{table}

\subsection{Temporal Dislocation}
\begin{figure}[ht]
    \centering
    \includegraphics[width=0.9\columnwidth]{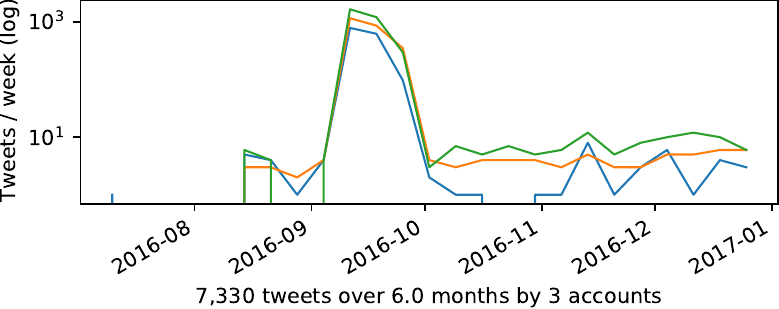}
    \caption{Weekly activity timelines for an HCC of IRA accounts in 2016. Note that the blue actor was inactive for several weeks. } 
    \label{fig:ira_hcc13_timelines}
\end{figure}

In the DS2 dataset, which covers a year of Twitter activity, using results from \cite{weber2021coord}, we plotted the retweeting activity of HCC members and identified at least one HCC where individual member activity patterns differed significantly (Figure~\ref{fig:ira_hcc13_timelines}). Though the members' activity aligns closely between September and October, after that the blue timeline deviates from the green and orange. This shows that if the collection had been restricted to the last two months of $2016$, the blue member may not have been included in the HCC.

\subsection{Cheerleaders}
We searched the RNC dataset for co-retweeting within $10$ second windows and then retained only edges in the resulting CN with a weight greater than $10$ (i.e., $>10$ instances of co-retweeting) and their adjacent nodes. The short timeframe was designed to identify \emph{social bots}, automated accounts presenting as human users used in astroturfing campaigns~\cite{Cresci2020,rise2016}.

\begin{figure}[th]
    \centering
    \includegraphics[width=0.8\columnwidth]{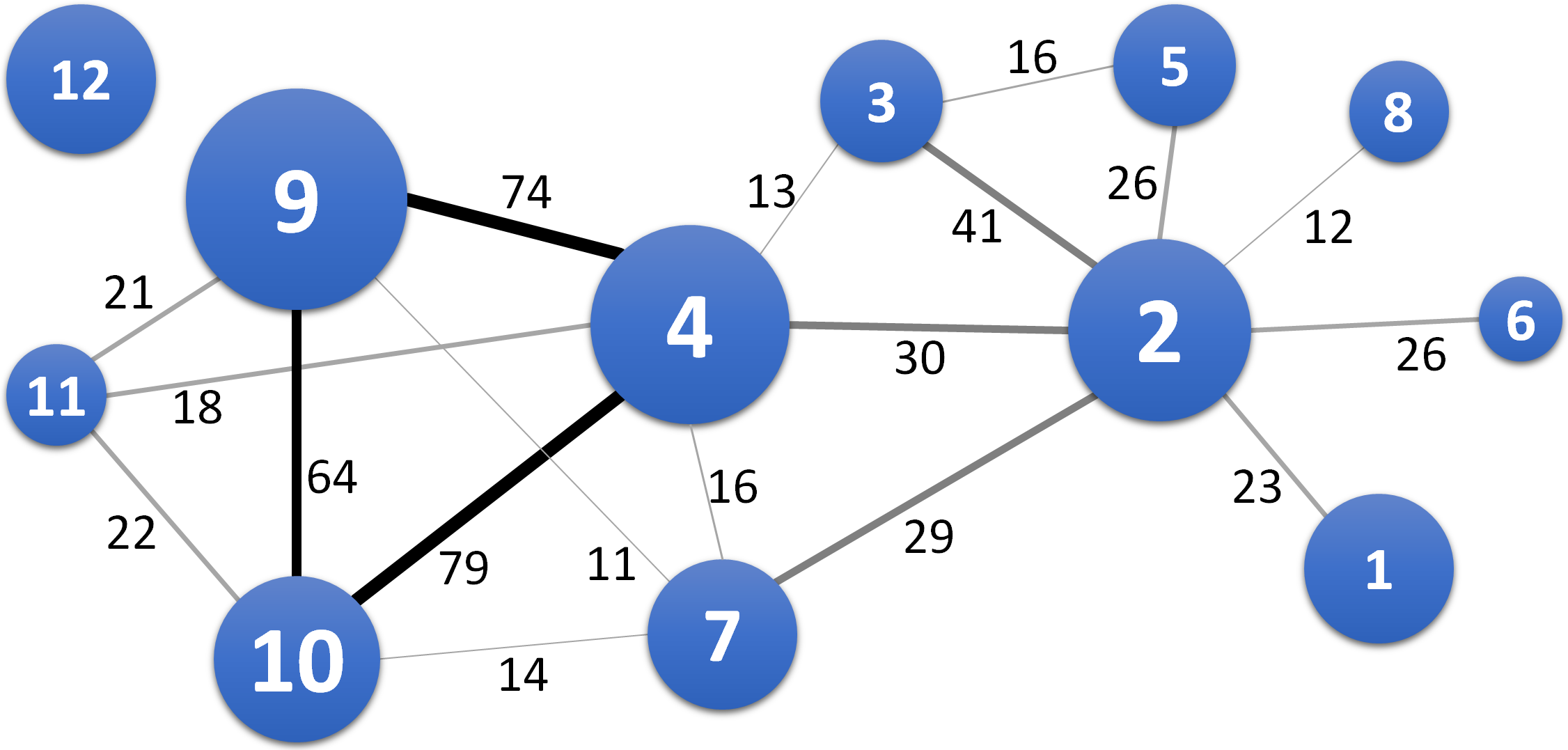}
    \caption{The co-retweeting HCC supporting \mention{TWR} (labeled $4$) using $\gamma$=$10$s and edge weight above $10$. Edge weights, darkness and thickness are the number of co-retweets. Node size indicates tweet activity and node labels are anonymised IDs. }
    \label{fig:twr_hcc}
\end{figure}

We identified a clear HCC supporting the official \mention{TrumpWarRoom} (\mention{TWR}) account (labeled $4$ in Figure~\ref{fig:twr_hcc}). Accounts $2$, $9$, and $10$ coordinate directly with \mention{TWR} $30$, $74$, and $18$ times, respectively, but they are all very active, suggesting their coordination may be a symptom of being very active accounts (in an agenda-focused community). They each contributed more than $1,000$ tweets overall including more than $200$ co-retweets, and one ($2$) forms the core of another star formation. Accounts $9$ and $10$ co-retweet $64$ times, each going first half the time, suggesting genuine coordination.

\begin{figure}[th]
    \centering
    \includegraphics[width=0.8\columnwidth]{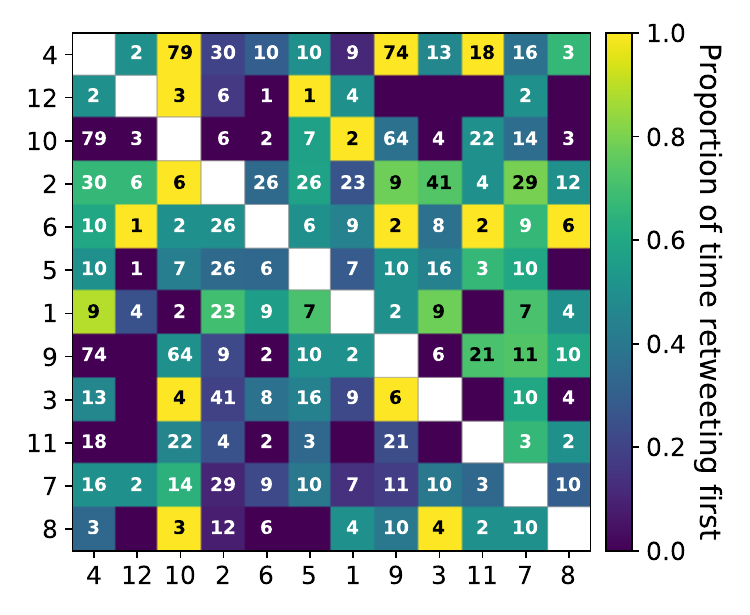}
    \caption{Number of times \mention{TWR} HCC accounts co-retweeted first when coordinating (i.e., within $\gamma$). Blank (blue) cells imply no co-retweeting between accounts. Cell colour indicates proportion of times the row account was first. E.g., \mention{TWR} ($4$) always co-retweets first with $9$, $10$, and $11$ but mostly second with $1$ and $2$. }
    \label{fig:twr_hcc_cheerleaders}
\end{figure}

To search for cheerleaders, we analyse the relative timings of co-retweets for each pair of HCC members, resulting in the grid in Figure~\ref{fig:twr_hcc_cheerleaders}. To co-retweet, two accounts each retweet the same tweet. Each square in the grid indicates how often the pair (denoted by row and column labels) co-retweeted, and the colour indicates how often the row account posted first. Account $4$ (\mention{TWR}) retweets first w.r.t. accounts $9$, $10$ and $11$, on every one of $74$, $79$ and $18$ times, respectively, which is striking for a $4$-day period. 
This visualisation, coupled with the edge information, shows that $9$ and $10$ are probably cheerleaders, and $11$ may be a cheerleader for \mention{TWR}, $9$ and $10$, and \mention{TWR} has a close relationship with $2$.

\begin{figure}[th]
    \centering
    \includegraphics[width=0.99\columnwidth]{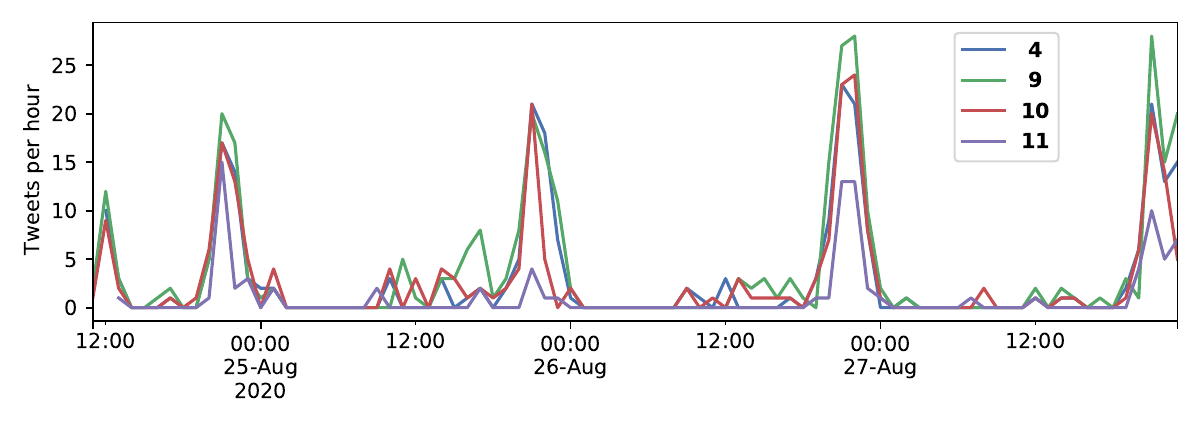}
    \caption{Hourly posting activity for the accounts $4$ (\mention{TWR}), $9$, $10$, and $11$. }
    \label{fig:twr_hcc_subset_timelines}
\end{figure}

Further evidence in Table~\ref{tab:cheerleader_deets} suggests that the accounts most strongly coordinating with \mention{TWR} are likely to be social bots. They are mostly long-lived accounts that have amassed considerable numbers of followers and post at very high rates (up to $92$ tweets a day for six years), but may have avoided Twitter's bot detection systems by maintaining reputation scores, where
\begin{equation}
    reputation = \frac{|followers|}{|friends| + |followers|}.
\end{equation}
Botometer Complete Automation Probability (CAP, English variant) \cite{davis2016botornot} ratings, however, suggest a very high chance the accounts are automated. 
Two accounts ($9$ and $11$) are well-established and have posted an average of over $30$ tweets a day for more than eight years, while the last ($10$) is young, but twice as active. The older accounts describe themselves as \hashtag{MAGA} supporters, so their support for \mention{TWR} is unsurprising. All co-retweeted second to \mention{TWR} within $10$ seconds over the convention, $171$ times together. Examining their posting activity over the RNC, it appears they tweeted at similar rates during the evenings, apart from $11$, which was less active, except on the first night (Figure~\ref{fig:twr_hcc_subset_timelines}). 
Foreign-driven bots and other inauthentic accounts supporting political candidates as part of covert foreign policy has been previously observed \cite{weisburd2016trolling}, but it is unclear whether this is an example of such.

\begin{table}[th]
    \centering
    \caption{Details of potential cheerleader accounts.}
    \label{tab:cheerleader_deets}
    \resizebox{\columnwidth}{!}{%
        \begin{tabular}{@{}crrrrrrr@{}}
            \toprule
            Account & Tweets & Age (yrs) & Tweets/day & CAP  & Friends  & Followers & Reputation \\
            \cmidrule{1-1} \cmidrule(l){2-4} \cmidrule(l){5-5} \cmidrule(l){6-8}
            2       & 213.4k & 6.4       & 92.1       & 0.73 & 1.8k  &  3.6k & 0.67 \\
            9       & 103.3k & 8.8       & 32.1       & 0.80 & 2.4k  &  1.7k & 0.42 \\
            10      &  24.5k & 0.9       & 78.7       & 0.79 & 24    &  118  & 0.83 \\
            11      & 111.7k & 8.5       & 36.1       & 0.92 & 12.3k & 11.3k & 0.48 \\
            \bottomrule
        \end{tabular}
    } 
\end{table}

These examples show that taking into account the temporal aspects of both co-actions and activity at the HCC level, helps to  develop a clearer picture of how genuine the coordination shown in CNs is, and to what extent accounts cheerlead for each other.

These methods can be developed further by also considering general account activity rather than just relying on frequency of co-actions, and by examining two-level account/reason CNs as described in \cite{weber2021coord}.

\section{Conclusion}\label{sec:conclusion}

In recent years, detecting coordinated behaviour between social media accounts has become a focus of research using a variety of methods, many exploiting network analysis. Although some methods attend to the temporal nature of the data under examination, none have explicitly considered the issue of implying associations between members of the same communities, when their temporal activity explicitly distinguishes them. This is a particular risk exemplified by cheerleader accounts that coordinate with others without their knowledge. Given the ongoing presence of online influence campaigns relying on both overt and covert coordinated behaviour, there is a clear need to develop more sophisticated detection solutions, and better understand the semantics of the coordination networks they create. 

\bibliographystyle{abbrv}
\bibliography{main}

\end{document}